**In-situ non-equilibrium nanomechanics in a proton-conducting ceramic at low temperatures**

*Oleg Yu. Gorobtsov\*, Yumeng Song, Kevin Fritz, Daniel Weinstock, Yifei Sun, Dina Sheyfer, Wonsuk Cha, Jin Suntivich, Andrej Singer\**

O. Yu. Gorobtsov, Y. Song, K. Fritz, D. Weinstock, Y. Sun, J. Suntivich, A. Singer
Department of Materials Science and Engineering, Cornell University, Ithaca, NY 14853, USA.
E-mail: gorobtsov@cornell.edu, asinger@cornell.edu

D. Sheyfer, W. Cha
X-ray Science Division, Advanced Photon Source, Argonne National Laboratory,
Lemont, IL 60439, USA.

**Abstract**: Nanostructured proton-conducting ceramics (PCCs) have attracted considerable interest as moderate-temperature proton conductors. Structure dynamics during proton conduction, particularly at grain boundaries, are crucial for stability and proton transport in nanostructured PCCs. A common assumption is that PCCs are structurally stable at low operating temperatures; however, material polycrystallinity, absorption, and reactive operating conditions have so far prevented verifying this assumption by nano resolved in-situ structure measurements. Here, in an archetypal PCC $BaZr_{0.8}Y_{0.2}O_{3-d}$ the premise of structural stability is demonstrated to be inaccurate at temperatures as low as 200 °C. Coherent X-ray diffraction on a nanostructured $BaZr_{0.8}Y_{0.2}O_{3-d}$ sintered pellet is adapted to image in-situ three-dimensional structural processes inside the constituent submicron grains in a humid nitrogen atmosphere at 200 °C. Direct observation reveals non-equilibrium defect generation and subsequent grain cracking on a timescale of hours, forming new, otherwise energetically unfavorable facets in $BaZr_{0.8}Y_{0.2}O_{3-d}$. Furthermore, the structural rearrangements correlate with dynamic inhomogeneities of the lattice constant within the grains, showing potential heterogeneous $H^+$ transport. Our results elucidate the mechanisms behind PCCs structural degradation, overturn existing assumptions about the structure dynamics in PCCs, and fill a method gap for further in-depth in-situ studies of the PCC nanostructure.

# 1. Introduction

The drive for energy-efficient electrochemical devices has brought much attention to proton-conducting ceramics (PCCs).[1,2] PCCs are unique solid electrolytes that acquire protons from ambient hydrogen and water vapor through equilibration with oxide lattice defects.[2,3] The efficient proton transport in PCCs has allowed solid-state electrochemistry at temperatures below 600 °C, making PCC-based devices attractive for electrochemical energy conversion and electrochemical manufacturing. To date, PCCs have shown promise for a wide range of technological applications, such as fuel cells for energy conversion[4,5,6] and membrane reactors for hydrogen production.[7,8] However, further advances in the performance and reliability of PCCs are still necessary,[9,10] in particular, to determine the long-term operation and microstructural stability of PCC-based materials.

Among the archetypal and most technologically relevant PCCs is the acceptor-doped barium zirconate,[11] particularly the yttrium-doped barium zirconate $BaZr_{0.8}Y_{0.2}O_{3-d}$ (BZY). Acceptor-doped barium zirconate is a highly chemically stable perovskite oxide with excellent proton conductivity below 600 °C, making it a popular candidate for fuel cells, electrolyzers, and electrochemical syntheses.[8,12,13,14] In the last decade, nanostructured PCCs composed of submicron grains have attracted interest due to their increased proton conductivity and lowered operative temperatures down to 400 °C.[6,15,16,17,18] In particular, the large interfacial area between grains can stabilize the interfacial hydrated layer to provide a pathway for protonic conduction. Nanostructured BZY, as an archetypal material, has also demonstrated excellent proton conductivity below 400 °C and high chemical stability.[18]

While a number of researchers have examined chemical processes in BZY,[12,19,20,21] the lack of in-situ structure characterization conceals the structure dynamics and evolution in PCCs, especially at the nanoscale, where the structure dynamics and interfacial processes play a critical role. In the bulk and film PCCs, the structural instability during operation can affect the performance of the PCC-based devices, as crystal structure modification can affect proton conduction.[9] In the BZY-based materials, strain and misfit dislocations demonstrably affect electrochemical performance.[22,23] The effects of nanostructure in BZY on the functional properties therefore cannot be ignored, but the nanostructure is generally assumed to be relatively static at the operative temperatures and timescale of hours. However, intra- and inter-particle stresses developed from chemical processes and the proton transport mean that the submicron-sized grains must eventually develop defects, strain gradients, and eventually cracks, changing the macroscopic properties.

The main challenge to answering whether and how the PCC nanostructure evolves in-situ resides in developing in-situ experimental methods to characterize the nanostructure. To understand the dynamics of structure and grain boundaries, one must extract information about strain, defects, and crystal coherence within single grains of submicron size embedded in a polycrystalline, 10-1000 micron thick pellet. The unique problem is to achieve a sub-100 nm resolution on the 3D defect imaging of a single grain surrounded by millions of similar grains, despite high absorption in a thick pellet and reactive operating conditions (humidity and temperature above 100 °C). It is challenging to image the evolving structure with sufficient resolution in-situ with electron microscopy due to the high absorption in the pellet and reactive operating conditions.[2] X-ray[24] and neutron[25] diffraction methods, in comparison, provide sufficient penetration depth for in-situ structure investigations in ceramic materials.[2] However, conventional X-ray and neutron diffraction only provides averaged information on the structure over multiple grains and inadequate spatial resolution.

Recently, Bragg coherent X-ray diffraction[26,27,28,29] has enabled imaging operando structure of individual grains in battery electrodes, where similar limitations exist. Coherent scattering from individual grains produces speckle patterns uniquely dependent on the internal structure and shape of the grain (example in **Figure 1**, a). Phase retrieval[29,30] on a 3D Bragg peak collected by rocking the sample (Figure 1, b) provides the 3D structure of the grains and the atomic displacement within (Figure 1, c), reaching sub-100 nm resolution for strain and particle shape and detecting non-equilibrium defects such as dislocations and domain boundaries.

Here, we adapt recently developed grain Bragg Coherent X-ray Diffractive Imaging (gBCDI)[31] to track in-situ the evolution of structure and defects within the individual grains in the $BaZr_{0.8}Y_{0.2}O_{3-d}$ pellet. In a polycrystalline material, gBCDI enables nanoscale imaging of the changes in atomic displacement and the evolution of defects in non-isolated submicron-sized grains, going far beyond incoherent X-ray diffraction capabilities. Combining in-situ diffraction analysis and gBCDI, we track the structure evolution within an archetypal nanostructured PCC material on an individual grain level in a humid environment at 200 °C. We find that, contrary to expectations, multiple non-equilibrium structural defects and new grain facets develop on a timescale of hours even at this temperature, which is in the low range of operating temperatures for electrochemical devices using BZY. Our results demonstrate that the evolution of structure and grain cracking is more prevalent in PCC materials than previously thought. Furthermore, we expect in-situ gBCDI measurements to fill a niche among in-situ studies on structural evolution in nanostructured PCC materials.

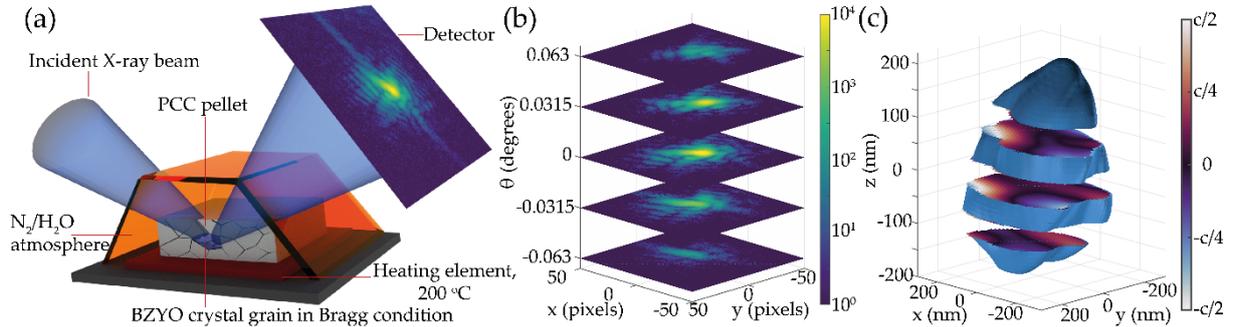

**Figure 1. Scheme of the BCDI experiment. a,** Focused incident coherent X-ray beam scatters from a single grain in the Bragg condition in a nanostructured ceramic pellet, producing a speckled diffraction pattern on the detector. The heating element maintains an increased temperature of the pellet in a humid $N_2/H_2O$ atmosphere. The pellet is rotated in the beam, producing a reciprocal space map of the diffraction peak, **b**. The distribution of intensity in the coherent diffraction peak contains the structural information to retrieve the 3D particle shape and the displacement field, **c**.

## 2. Results and Discussion

### 2.1. Experimental setup and pre-characterization of the $BaZr_{0.8}Y_{0.2}O_{3-d}$ pellet

We prepared the $BaZr_{0.8}Y_{0.2}O_{3-d}$ (BZY) pellets from crystalline BZY powders (**Figure 2**, a), which we first formed from nitrate precursors via a sol-gel synthesis followed by calcination at 900 °C for 5 hours in the air (3 °C $min^{-1}$ heating rate). Subsequently, we pressed the crystalline powders into ~50 mm diameter pellets and heated them at 1100 °C for 36 h in the air (1 °C $min^{-1}$ heating and cooling rate) to sinter the grains. Laboratory x-ray diffraction (XRD) patterns for powder and pellet can be indexed to $BaZrO_3$ (matching International Centre for Diffraction Data PDF 01-089-2486) (Figure 2, a). SEM images indicate that the pellet has significant porosity, consistent with the low sintering temperature (Figure 2, b).

We performed the in-situ coherent x-ray measurement at the beamline 34 ID-C of the Advanced Photon Source (Argonne National Laboratory, ANL, USA). The photon energy was 9 keV, focus size 400x400 nm, with the sample-detector distance of 1 m, and ASI Quad (512x512) Timepix detector with 55x55 μm pixel size. We collected (110) Bragg diffraction peaks (scattering angle 26.6 degrees) from several individual grains in the sintered BZY pellet for over 30 hours at 200 °C in a humid nitrogen atmosphere (setup scheme in Figure 1, a). Collecting full 3D reciprocal space maps of the Bragg diffraction peaks requires 1-3 minutes of rocking the sample chamber in the scattering plane (schematically shown in Figure 1, b). A full

Bragg peak angular spread is below 1 degree, and an angular step below 0.01 degree is required to sufficiently oversample the speckle pattern for phase retrieval.[30] Bragg diffraction peaks from individual grains remained stable over hours in a pure nitrogen atmosphere at 200 °C without introducing humidity, therefore excluding significant radiation damage effects on the in-situ measurement.

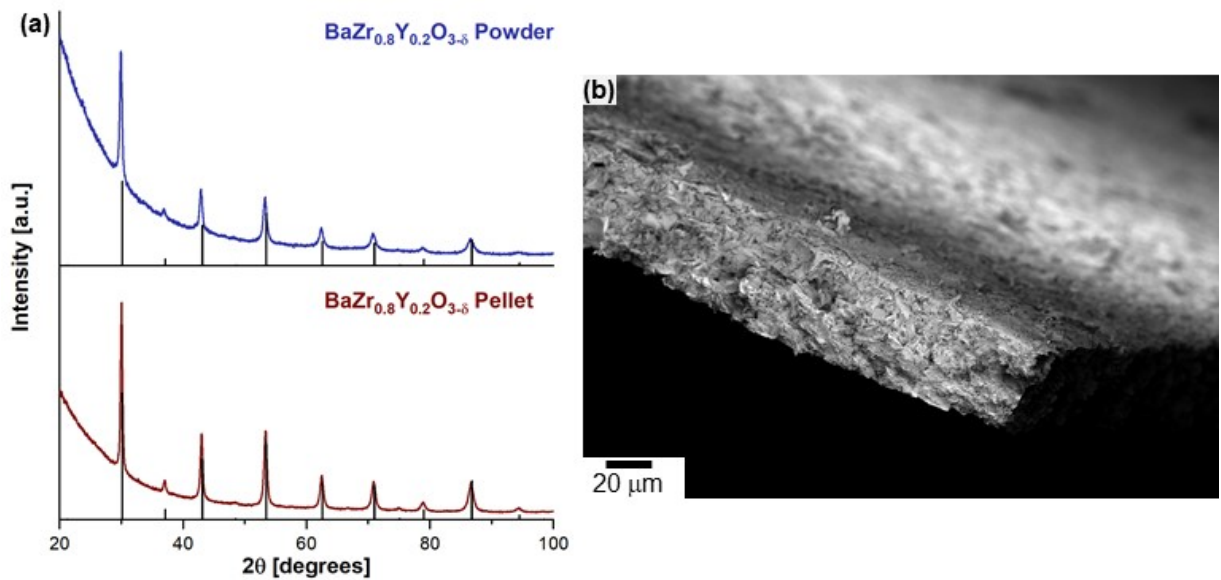

**Figure 2. Pellet characterization. a,** Lab-scale XRD pattern for $BaZr_{0.8}Y_{0.2}O_{3-d}$ powder and $BaZr_{0.8}Y_{0.2}O_{3-d}$ pellet matches with PDF 01-089-2486, **b,** Cross-sectional SEM image of $BaZr_{0.8}Y_{0.2}O_{3-d}$ pellet

## 2.2. Direct analysis of Bragg coherent x-ray scattering

Analysis of the reciprocal space maps from grains within the same $BaZr_{0.8}Y_{0.2}O_{3-d}$ pellet provides immediate information on the comparative structural evolution of multiple grains (**Figure 3**). Even without the real-space imaging with phase retrieval (discussed later), significant changes in the diffraction patterns are noticeable by the eye. Unexpectedly, we find a slow, on a timescale of hours, split and separation of single Bragg diffraction peaks into multiple peaks (Figure 3, a). The angular separation between the splitting peaks grows initially with a speed of approximately ~0.5 mrad/hour, before the separation rapidly increases. At this point, the diffraction peaks separate further away than the angular detector size and are not trackable simultaneously. Splitting occurs mainly perpendicular to the scattering vector **q**. The remaining total scattered intensity in the brighter diffraction peak after complete separation is 2-3 times smaller than the intensity before the split, showing a steep decrease in the crystal coherent volume within the grain (Figure 3, b, grains P4 and P5). The peak splitting suggests that the crystal grain splits initially into slightly misaligned domains, producing separated by ~1 mrad scale angle, but still simultaneously visible peaks. Subsequently, growing misorientation suggests the fracture of a single grain into two different grains of a smaller individual volume. Because diffraction of both domains is visible while illuminated with a ~1 μm x-ray beam, the domains likely remain close. The peak splitting is due to relative angular misorientation between the domains. The slow (over several hours) speed of the misalignment is likely due to restrictions imposed by the neighboring grains.

To further investigate the structural deformation during fracture, we investigate the peak widths during and after crack propagation. Interestingly, the Bragg peak width both along and perpendicular to the scattering vector **q** (Figure 3, c) does not demonstrate a preferred increase or decrease of the width over the different grains. Degradation of the crystal structure commonly presents itself in the growing average strain and number of defects in the grains, increasing the Bragg peak width. However, while particles 4 and 5 in (Figure 3, c) present a clear tendency to increasing peak width up to 40% parallel to **q** (thus not caused simply by peak splitting, which happens perpendicular to **q**), signifying increase in strain and/or defects, the peak width decreases rapidly after the peaks entirely separate. The decrease in peak width suggests that the stress and non-equilibrium defects in the grain are relieved by grain separation (cracking) into multiple smaller crystals. The absence of lingering strain gradient suggests brittle fracture with no significant permanent structural rearrangements away from the crack surface. The rest of the particles present diverging behavior, with peak width variation within 10-20% higher or lower than the pristine state. It is important to note that peak width perpendicular to **q** is inaccessible by conventional XRD, in which diffraction structure over the direction perpendicular to **q** is averaged out, and thus insensitive to the peak splitting observed here.

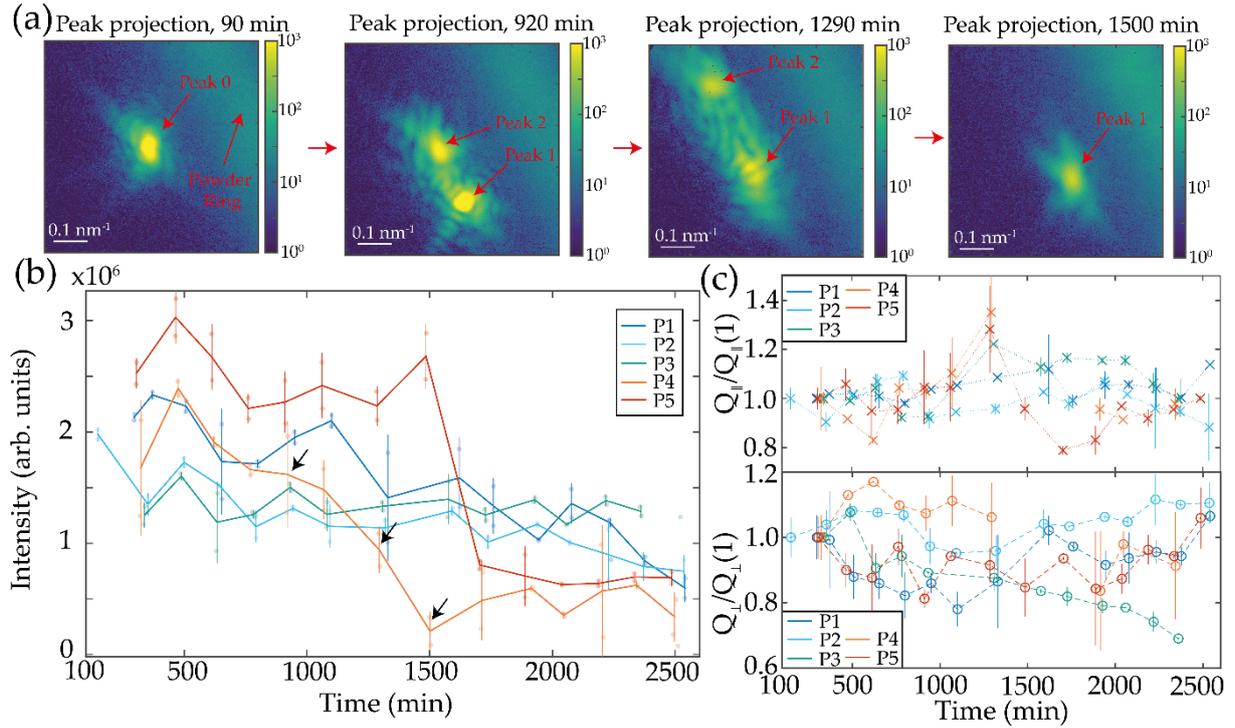

**Figure 3. Bragg Coherent X-ray Scattering. a,** Example of a Bragg peak splitting during the in-situ measurement. Grain P4 in **b** and **c** depicted. **b,** Total intensity of a Bragg peak for different grains. Error is calculated as a variation between two measurements at the same time. **c,** Evolution of the relative peak width parallel (top) and perpendicular (bottom) to the scattering vector for different grains.

## 2.3. Results of 3D coherent diffraction imaging

We further investigated the evolution of grains shape and internal structure by performing phase retrieval[29] on the collected Bragg diffraction peaks. We have successfully retrieved in three dimensions the shape of and the atomic displacement field within grains at specific times during the 30-hour period. Interestingly, even when the total scattering intensity decreases only by 10-20% and without apparent peak separation, as in grain P2 in Figure 3, b, a sharp change in the grain shape consistent with cracking is visible (example in **Figure 4**, a, top). The grain of approximately 500x500x500 nm size changes shape at approximately 2000 minutes into the in-situ measurement. Part of the volume present at 1000 – 1800 minutes, marked by a green circle in Figure 4, a, disappears at 2370 minutes and beyond, signifying the loss of crystal coherence with the rest of the grain. In the 3D coherent Bragg peak itself, the change is accompanied by a disappearance of a satellite maximum (marked by green arrows in Figure 4, a, bottom). Overlapping reconstructed grain shapes at 1590 min and at 2370 min (Figure 4, b) confirms the disappearance of a crystal volume. The comparison of the shape before and after the fracture allows us to determine the orientation of the crack plane in comparison to the scattering vector

**q**∥z, oriented normal to a crystallographic plane from {110} family (green). We find an angle of ~50-60 degrees, most closely matching to a plane from {112} family (brown). (112) crystallographic plane is oriented at an angle 54 degrees to (110) plane in the BZY perovskite crystal structure (Figure 4, c).

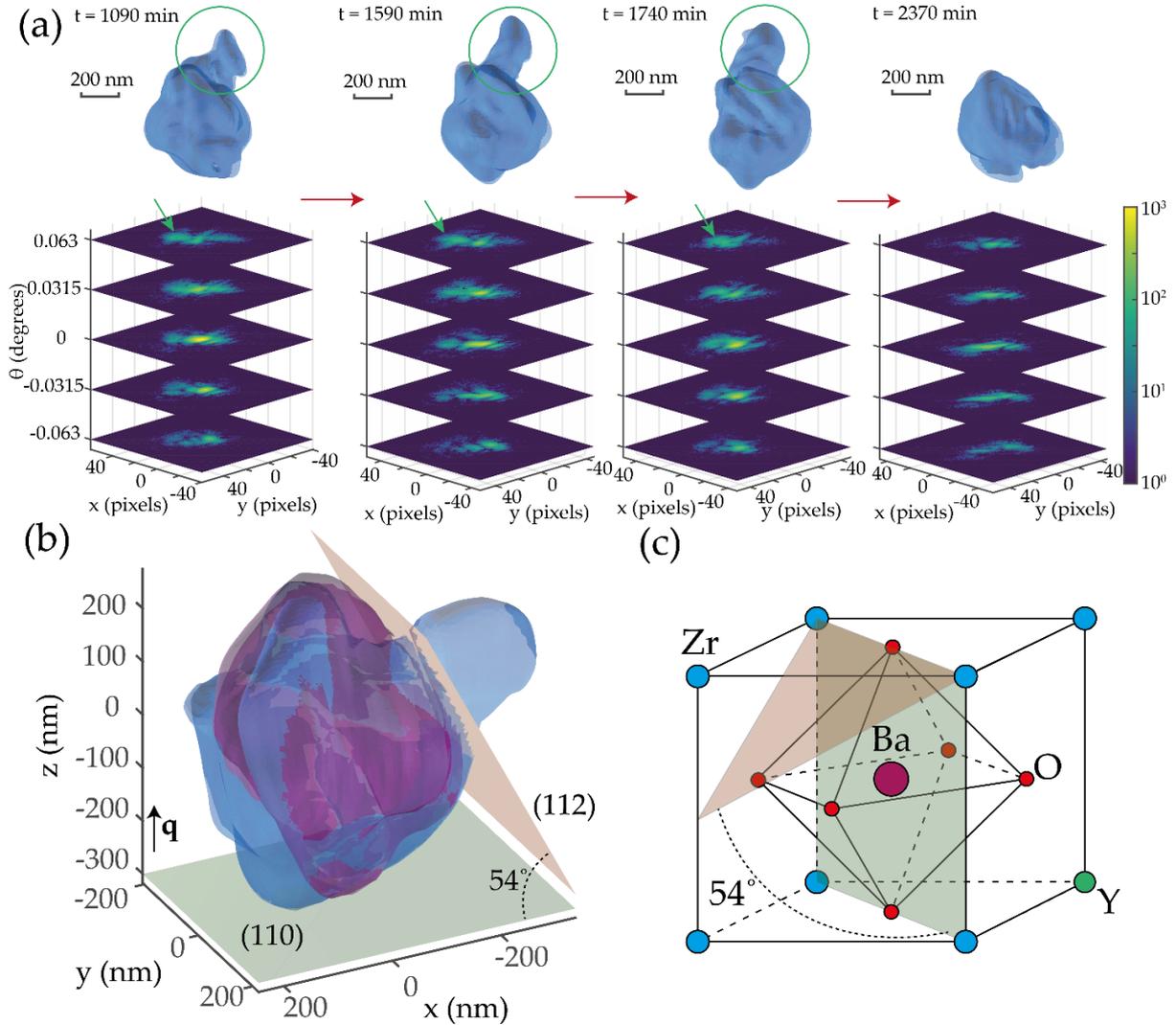

**Figure 4. Imaging results - particle cracking. a,** Example of the changes in the Bragg peak and the changes in shape for a reconstructed grain (P2 in Figure 3). Isosurface at 15% maximum amplitude, slight variation in shape is explained by the uncertainty in the modulus of the retrieved complex amplitude. **b,** Example of the cracking. Blue surface - particle shape at t=1590 min, magenta surface - particle shape at t=2370 min, green horizontal plane - (110) crystallographic plane, brown plane - (112) crystallographic plane. **c,** BaZr$_{0.8}$Y$_{0.2}$O$_{3-d}$ unit cell schematic with (110) and (112) planes marked.

In equilibrium, dominant facets of BZY crystals are along {001} and {110} plane families.[32,33,34] In BZY nanocrystals, {111} facets have been observed.[35] Therefore, cracking along the {112} plane family, leading to {112} facets, is unexpected. Previously, similarities between BZY and CeO$_2$ nanocrystals have been found,[35] and in CeO$_2$ {112} planes are possible termination planes,[36] although they spontaneously turn into a stepped {111} surface. The resolution of our measurement is insufficient to observe a surface rearrangement to a stepped surface; however, to the authors' knowledge, no {112} termination planes have been previously reported in BZY. Different lattice constants induced in the separating volumes before facet formation, suggesting different H$^+$ concentrations, lead us to speculate, therefore, that the non-equilibrium effects and the interaction with the neighboring grains make {112} termination plane energetically more favorable.

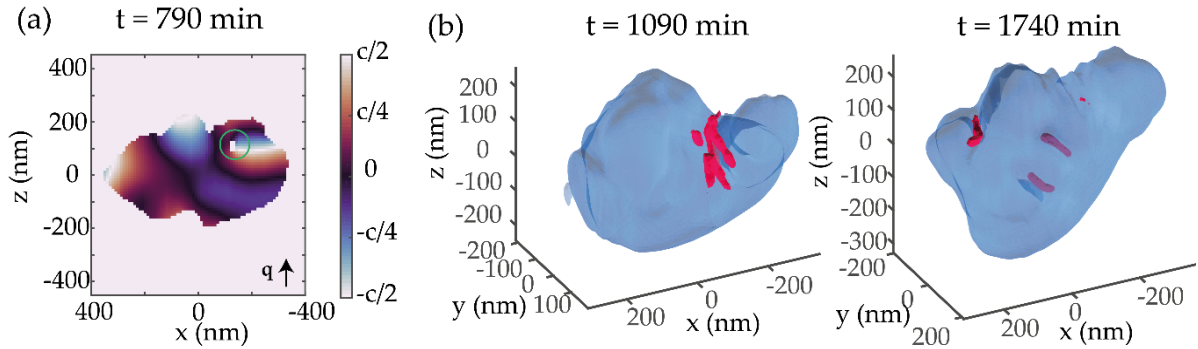

**Figure 5. Typical dislocations in the grain. a,** Sample cross-section of a reconstructed displacement inside a grain, showing a vortex and a drop in amplitude (green circle) signifying a dislocation. In 3D, the dislocation line (red) goes through the grain (blue). **b,** Dislocation lines (red) in the grain P2 (blue) on the border of the cracking region. Note that the surface of the grain is imperfectly defined due to the number of dislocations introducing zero values in the amplitude.

Furthermore, the complex phase of the real space complex amplitude retrieved through gBCDI provides in-situ information on the 3D distribution of atomic displacement within the grains in the [110] crystallographic direction. Analysis of the atomic displacement within the P2 grain demonstrates the abundance of dislocations generated during the in-situ process (**Figure 5**). Dislocations with a component of the Burgers vector b along the scattering vector q produce a singularity in the atomic displacement. They can be pinpointed as vortices in the displacement field (Figure 5, a, marked by a green circle), also producing zeroes ("holes") in the reconstructed shape (see the center of the vortex in Figure 5, a) because of undefined displacement at the dislocation core. We pinpoint the dislocation lines in 3D (Figure 5 b, red lines) by tracking the singularities in the retrieved displacement through the grain. Multiple dislocations with different orientations of the dislocation line are found in the grain P2, evolving over time. Interestingly, the grain volume that later detaches demonstrates a particular proclivity for dislocations (Figure 5, b). Note the jagged appearance of the grain surface in the region due to the zeroes in amplitude produced by dislocations. While the orientation of the dislocation lines differs, all of them have a component in the (110) plane, perpendicular to the scattering vector q. Note that a screw dislocation with a dislocation line entirely in the (110) plane would not produce a vortex in the atomic displacement because the Burgers vector would be oriented perpendicular to the scattering vector q||[110], which suggests that the dislocations are preferentially of the edge type. Additionally, our experimental geometry is only sensitive to dislocations with the Burgers vector not perpendicular to the Q vector, suggesting there might be more dislocations we do not see in the displacement field.

While perovskites do not form an isomechanical group, in perovskites such as $SrTiO_3$ and $KNbO_3$, and theoretically generally in perovskite oxides, edge dislocations aligned along <110> at low temperatures (<1000 K) are mobile and dissociate producing stacking faults.[37,38] Our in-situ imaging results show that the dislocation configuration changes at a sub-hour timescale in BZY, showing experimentally similar <110> dislocation behavior to the one theoretically predicted for other oxide perovskites.

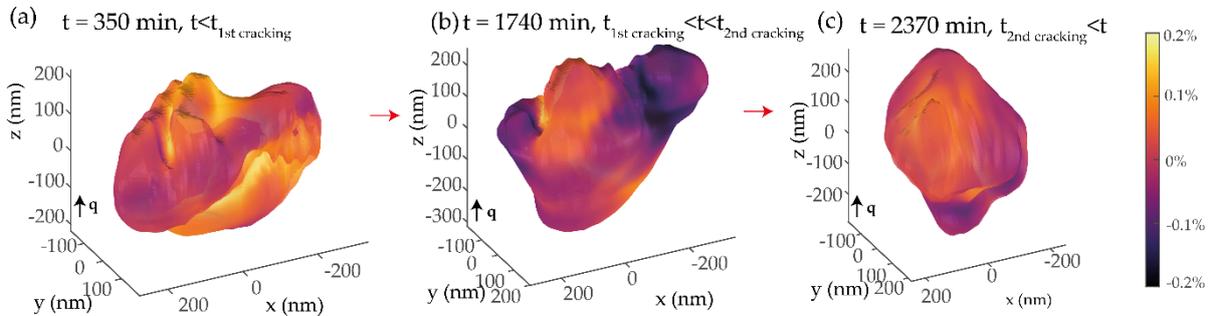

**Figure 6. Spatially resolved strain evolution. a,** Distribution of strain in the pristine grain (surface at 15% amplitude). **b,** Distribution of strain in the grain before the cracking event depicted in Figure 4 (surface at 15% amplitude). **c,** Distribution of strain in the grain after cracking (surface at 15% amplitude).

The displacement field provides information about the distribution of the strain in the [110] direction, which is a derivative of the dislocation field along the scattering vector. Analysis of the strain distribution (**Figure 6**) shows a significant spatial difference in strain accumulation across the grain. In the beginning stages of the process, the strain is distributed homogeneously (Figure 6, a), with a variation of +-0.1% of the crystal lattice spacing. However, after the first ~1500 minutes, the accumulated strain in the main and detaching volumes of the grain differ by approximately 0.4%. More precisely, the average lattice spacing in the volume that detaches after the cracking event seen in Figure 4 is 0.4% lower, signifying either evolving external stress from neighboring grains or a lower penetration by $H^+$ ions. Note that the strain difference between the center of the grains and their surface is, in comparison, much smaller (<0.1%), suggesting a more homogeneous ion distribution within the two volumes. Taken together, our result suggests that BZY undergoes cracking in humified environment even in absence of applied electrochemical potential and under mild heating (200 °C). This finding indicates the importance of fracture toughness in PCCs and suggests that some of the active BZY materials are likely lost and thus cannot participate in electrochemistry during temperature and humidity cycling, for example, during the start-up and shut-down cycles. Finding PCCs with superior fracture properties is, therefore, essential to ensure the longevity of moderate-temperature electrochemical energy devices.

## 3. Conclusion

In summary, we reveal the in-situ behavior of nanostructure in PCCs by adapting gBCDI to image the structural dynamics in a nanostructured $BaZr_{0.8}Y_{0.2}O_3$ pellet. We find unexpected structural activity at 200 °C in a humid nitrogen atmosphere, specifically, cracking of the grains and the abundance of mobile dislocations that align preferentially along the {110} plane. We have imaged cracking of the grains along the {211} crystallographic planes, which generates facets energetically unfavorable in equilibrium conditions. The crack occurs in the vicinity of the dislocations, suggesting strong interaction between defects. Additionally, we found the formation of clearly distinct regions with different lattice constants correlated with cracking. Given that most PCC devices operate between 300 – 600 °C, the observed microstructural evolution and grain instability of PCCs at low temperature (200 °C) and in the absence of electrochemical reaction merits further investigation of structural degradation in PCCs. Our results suggest the potential use of gBCDI as a tool for screening the mechanical properties of future nanostructured PCC candidates in-situ and operando in electrochemical devices.

**Acknowledgments**

The work at Cornell was supported by the National Science Foundation under Grand No. (CAREER DMR 1944907). J.S. acknowledges support from the NSF CHE 1665305 Project. This research used resources of the Advanced Photon Source, a U.S. Department of Energy (DOE) Office of Science User Facility, operated for the DOE Office of Science by Argonne National Laboratory under Contract No. DE-AC02- 06CH11357.

**Competing interests**

The authors declare no competing interests.